\documentclass[fleqn,usenatbib]{mnras}

\usepackage[T1]{fontenc}

\DeclareRobustCommand{\VAN}[3]{#2}
\let\VANthebibliography\thebibliography
\def\thebibliography{\DeclareRobustCommand{\VAN}[3]{##3}\VANthebibliography}

\usepackage{graphicx}	% Including figure files
\usepackage{amsmath}	% Advanced maths commands
\usepackage{amssymb}	% Extra maths symbols
\usepackage{amsbsy,amsfonts}

\usepackage{comment}
\usepackage{psfrag}
\usepackage{hyperref,graphicx}    

\newcommand*{\figref}[2][]{%
  \hyperref[{#2}]{%
    Figure~\ref*{#2}%
    \ifx\\#1\\%
    \else
      \,#1%
    \fi
  }%
}
\newtheorem{theorem}{Theorem}
\newtheorem{corollary}{Corollary}

\usepackage{tikz}
\usetikzlibrary{arrows,matrix}
\newcommand{\opm}{
        \mathbin{
                \mathchoice
                {\buildcirclepm{\displaystyle     }{0.14ex}{0.95}{0.05ex}{.7}}
                {\buildcirclepm{\textstyle        }{0.14ex}{0.95}{0.05ex}{.7}}
                {\buildcirclepm{\scriptstyle      }{0.13ex}{0.955}{0.05ex}{.55}}
                {\buildcirclepm{\scriptscriptstyle}{0.08ex}{0.95}{0.03ex}{.45}}
                }
                }

\newcommand\buildcirclepm[5]{%
        \begin{tikzpicture}[baseline=(X.base),inner sep=-#5, outer sep=-.75]
                \node[draw,circle,line width=#4] (X) {\footnotesize\raisebox{#2}{\scalebox{#3}{$#1\pm$}}};
        \end{tikzpicture}%
}

\title[Analytical 1D solutions of a working surface]{Full analytical
ultra-relativistic 1D solutions of a planar working surface}

\author[M.E de la Cruz-Hern\'andez \& S. Mendoza]{
Manuel E. de la Cruz-Hern\'andez\thanks{E-mail:
mdelacruz@ciencias.unam.mx (MEC)}
and Sergio Mendoza\thanks{
E-mail: sergio@astro.unam.mx (SM)}
\\
Instituto de Astronom\'{\i}a, Universidad Nacional
                 Aut\'onoma de M\'exico, AP 70-264, Ciudad de M\'exico 04510,
                 M\'exico
}

\date{Accepted XXX. Received YYY; in original form ZZZ}

\pubyear{2021}

\begin{document}
\label{firstpage}
\pagerange{\pageref{firstpage}--\pageref{lastpage}}
\maketitle

\begin{abstract}
  We show that the 1D planar ultra-relativistic shock tube problem
with an ultra-relativistic polytropic equation of state can be solved
analytically for the case of a working surface, i.e. for the case
when an initial discontinuity on the hydrodynamical quantities of the
problem form two shock waves separating from a contact discontinuity.
The procedure is based on the extensive use of the Taub jump conditions
for relativistic shock waves, the Taub adiabatic and performing Lorentz
transformations to present the solution in a system of reference adequate
for an external observer at rest.  The solutions are found using a set of
very useful theorems related to the Lorentz factors when transforming
between systems of reference.  The energy dissipated inside the 
working surface is relevant for studies of light curves observed in
relativistic astrophysical jets and so, we provide a full analytical
solution for this phenomenon assuming an ultra-relativistic periodic
velocity injected at the base of the jet.
\end{abstract}


%%%%%%%%%%%%%%%%%%%%%%%%%%%%%%%%%%%%%%%%%%%%%%%%%%

%%%%%%%%%%%%%%%%% BODY OF PAPER %%%%%%%%%%%%%%%%%%

%%%%%%%%%%%%%%%%%%%%%%%%%%%%%%%%%%%%%%%%%%%%%%%%%%

\section{Introduction}
\label{introduction}

% Here mention jets, shocks, history of ballistic solutions and mention
% well why a working surface is formed.

  Relativistic astrophysical jets moving with speeds close to that of
light are common highly energetic phenomena in the Universe.  They cover
a wide range of luminosities, from values of $\approx 10^{47}$ erg
s$^{-1}$ in a microquasar, to  $ \gtrapprox 10^{50}$ erg s$^{-1}$
in long Gamma Ray Bursts (lGRBs),   with intermediate values of $\approx
(10^{42}-10^{47})$ erg s$^{-1}$ in Active Galaxy Nuclei (AGN)~\citep[see
e.g.][]{sunyaev2011,ghisellini2017}.  Their corresponding physical lengths
vary from \( \lesssim 1 \text{pc} \) for microquasars and GRB's to kpc
or a even a few Mpc for the case of AGN.  These highly collimated jets
emanate from neutron stars or black holes \citep{kulkarni}, and  are
surrounded by an accretion disc.

The interaction of these jets with inhomogeneities of the surrounding
medium, the deflection of the jets and the fluctuations in time of the
initial ejection parameters produce knots that are regions of increased
brightness, which can be interpreted as internal shock waves~\citep[see
e.g.][]{rees78,mendoza00,mendoza01,mendoza02}.

The space-time formation and propagation of a \emph{working surface},
i.e. two shock waves separating from each other from a contact
discontinuity in between them, has been used as a proposal that
determines the physical mechanism for the generation of internal shock
waves along a relativistic jet.   A working surface constitutes a
particular type of solution of the shock tube problem in hydrodynamics when 
a set of initial discontinuities appear on the
flow~\citep{landaufm,marti94,lora}.

  The semi-analytical approach to the formation of a working surface
presented in \citet{MHOC} assumes that the radiation
scale times are smaller than the characteristic dynamical times
of the problem and consequently the pressure of the fluid is
negligible. That approach further assumes time variations on the
ejected flow and its discharge.  That semi-analytical model has
been used to describe the mechanism of the generation of shock
waves in GRBs~\citep{MHOC}, microquasars~\citep{coronado2015} and
AGNs~\citep{Benitez2013,coronado2016}.  All these studies have consisted
in the statistical fit  of this semi-analytical model to the observed light
curves. The procedure provides best fit parameters associated with
the ejection mechanism.  Although successful, that semi-analytical
approach does not describe the flow in a full hydrodynamical manner,
since the fluid pressure is a fundamental parameter that contains
important information about the fluid.

  The pioneering work of \citet{marti94} to solve the relativistic shock
tube problem is directly related to the particular case that we solve in
the present article for the case of a relativistic working surface moving
through a relativistic jet, i.e. for the case of a pair of ultra-relativistic shock waves 
separated by a contact discontinuity that move through a hot\footnote{As it 
is commonly called in the relativistic high energy} astrophysical literature a hot (cold) gas is one for
which its internal energy density is much greater (smaller) than its 
rest-mass energy density.  This is not to be confused with the common definition
used in physical phenomena, where a hot gas is one for which its internal energy is greater 
than its kinetic energy, regardless the value of its rest-mass energy density.  external medium. The difference is that in the
present work we construct full exact analytic solutions of this particular
problem in the ultra-relativistic regime.

  For completeness we mention that in the context
of spherically symmetric working surfaces moving
through a cold medium, solutions were first 
constructed by \citet{katz94} and
further generalised by the works of \citet{piran94,sari95} (see also
\citet{piran-review-99,piran-review}).  \citet{katz94} solved the problem
in the particular system of reference of each individual shock wave.
\citet{piran94,sari95} used the general shock wave jump conditions of
\citet{mccolg73} to make the shock waves move to the adequate system of
reference of an external observer at rest.  Furthermore and in the same
spherically symmetric context, \citet{uhm2006} and~\citet{uhm2011} tried
different approaches to find analytical and semi-analytical solutions
of relativistic blast waves (blast meaning the gas in between both shock
waves --forward shock and reverse shock).  That analysis greatly differ
from the one we are presented since they conclude that the pressure
between both shocked gases have not the same value --as opposed to the
standard results of the shock tube problem and the discontinuities in
the initial hydrodynamical conditions that lead to the formation of a
working surface~\citep[e.g][]{landaufm,marti94}.  In
these works the authors tried to match the forward shock wave with the
self-similar blast shock wave of \citet{blandford76} but without using
the imploding self-similar shock wave solution of \citet{hidalgo05} for
the reverse shock wave.    In any case, all these spherically symmetric
solutions deviate from the purpose of the present work since they deal
with a working surface moving through a cold medium.

  In this article,  we present a full analytical model of an
ultra-relativistic working surface that moves through a
hot relativistic
medium.  The external and internal flows are assumed to obey a polytropic
ultra-relativistic equation of state.  This exact analytical solution of
the  working surface in relativistic hydrodynamics is an extension of
the full analytical non-relativistic hydrodynamical model presented
e.g.~by~\citet{landaufm}.  For astrophysical applications, the idea
is that the working surface is created from an initial discontinuity
generated by a variable injection velocity and a constant  density
discharge at the base of a relativistic astrophysical jet.

  The article is organised as follows.  Section~\ref{discontinuities}
presents basic relations of a general 1D planar relativistic flow
with an ultra-relativistic Bondi-Wheeler polytropic equation of state
and we describe in general terms the formation of a working surface.
In Section~\ref{rel-shock} we study the appropriate jump conditions
for a relativistic fluid and the formation of a working surface,
i.e.  two shock waves moving in the same direction and separated by a
contact discontinuity by an adequate choice of the system of reference.
In section~\ref{analytic} a full analytical solution is found for a one
dimensional planar working surface, valid for very strong shock waves
that move at ultra-relativistic velocities.  In Section~\ref{profiles},
the hydrodynamical profiles associated with the working surface are
obtained.  We show in this section the convergence of our solution
with known analytical and numerical results.  As an example, in
Section~\ref{workingsurface-energy} we generate a working surface by a
periodic injected velocity at the base of the relativistic jet, with the
assumption that the thermal energy inside the working surface is fully
radiated away by some efficient process. With this assumption we can
calculate the mechanical power, or radiated luminosity, generated by the
working surface.  As a result of all this, a the light curve emitted by
the working surface can be analytically calculated.  This light curve
is a function of of the hydrodynamical conditions of the flow inside
the jet, i.e. of the external flow to the working surface.

\section{Discontinuities on the initial conditions in relativistic
         hydrodynamics.}
\label{discontinuities}

  When a moving fluid reaches velocities close to that of light,
the equations describing its state and evolution must take into
account relativistic effects. The problem of shock wave propagation,
as an application of the methods of the theory of relativity to
problems in hydrodynamics is described by the continuity or mass
conservation equation, as well as the energy and momentum conservation
equations~\citep[e.g.][]{landaufm,mitchell}.

  For a perfect fluid, in the absence of force fields the shape of the
energy-momentum four-tensor $T^{\mu\nu}$ is given by

\begin{equation}
    T^{\nu \mu} = ( e + p )u^{\mu}u^{\nu} - 
      g^{\mu \nu}p.
\label{energy-momentum}
 \end{equation}

\noindent   In the previous equation, $u^{\mu}$ is the four-velocity of
a particular fluid particle measured from a given system of reference
and $g^{\mu \nu}$ is the metric tensor, which for a flat Minkowski
space-time in one spatial dimension takes the form  $g^{\mu \nu}=\mathrm{
diag}(1,-1,0,0)$.  All thermodynamical quantities such as the total energy
density (rest mass energy density $\rho$ plus internal energy density \(
\epsilon \)), the pressure $p$ and the enthalpy density \( w = e +
p \) are measured on their proper system of reference.  In here and in
what follows we adopt a metric signature (\(+,-,-,-\)) and an Einstein
summation convention is used, for which Greek indices take space-time
values \( 0, 1, 2, 3 \) and Latin ones take space values \( 1, 2, 3 \).
Here and in what follows, we use a system of units for which the speed
of light is $c=1$.

  For the case of planar symmetry in one dimension, the null
four-divergence of the energy momentum tensor~\eqref{energy-momentum}
takes the following form:

\begin{equation}
  \frac{\mathrm \partial T^{\beta\alpha}}{\mathrm \partial x^\beta} =
    \frac{\mathrm \partial T^{0\alpha}}{\mathrm \partial t} + \frac{\mathrm
    \partial T^{i\alpha}}{\mathrm \partial x^i} = 0.
\label{eq5.7}
\end{equation}

  The conservation of the proper particle number density $n$ is represented
by the equation of continuity: 

\begin{equation}
  \frac{\mathrm \partial \rho u^\alpha}{\mathrm \partial x^\alpha}
    =\frac{\partial \rho u^0}{\partial t}+\frac{\partial \rho u^k}{\partial
    x^k}=0.
\label{eq5.17}
\end{equation}

  To close the system of equations, we use a relativistic polytropic
equation of state given by~\citep{tooper65}:

\begin{equation}
 e=\rho +\frac{p}{\kappa-1},\label{eq7.8}
\end{equation}

\noindent where $\kappa$ is the polytropic index. In some astrophysical
situations when ultra-relativistic gases are present, the pressure $p$
is much greater than the rest mass density $\rho $ and the resulting
relationship is the Bondi-Wheeler \citep{bondi-64} equation of state:

\begin{equation}
  p=(\kappa-1)e.
\label{eq7.9}
\end{equation}

  For the case of a gas for which its constituent particles
(e.g.~electrons) move  at ultra-relativistic velocities, which are the 
cases we are interested in this work, the polytropic
index  $\kappa=4/3$~\citep{landauest}.

\begin{figure}
    \begin{center}
      %\psfrag{n}{$\bf{n} a$}
      \psfrag{vs1}{$v_\text{s(1)}$}
      \psfrag{vs2}{$v_\text{s(2)}$}
	\psfrag{v1}{$v_\text{(1)}$}
	\psfrag{v2}{$v_\text{(2)}$}
	\psfrag{S}{$S$}
	\psfrag{T}{$T$}
      \includegraphics[scale=0.45]{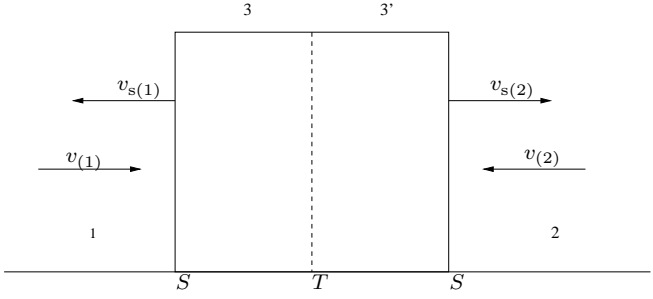}
    \end{center}
	\caption[Working surface generated by an initial
	discontinuity.]{The figure shows two shock waves $S$ separated
	by a stationary contact discontinuity $T$ with respect to the gas
	flow  surrounding it. The hydrodynamical system is generated from
	a particular set of initial discontinuities of the hydrodynamic
	quantities of the flow. This set of discontinuities and the flow
	between them is called a working surface. The arrows indicate the
	direction of the movement of the shock waves (continuous lines)
	and the movement of the flow.  The numbers label different
	regions separated by each discontinuity.}
\label{initial}
\end{figure}

One of the most important reasons for the presence of a working surface
in a gas flow is the possibility of discontinuities in the initial
conditions of the hydrodynamical quantities such as velocity, density
and pressure distributions. For example, suppose we have a flow with
variable velocity. If an element of the fluid ejected at an earlier
time has a lower velocity as compared to that of another fluid element
ejected at a subsequent time, then eventually rapid flow will reach
the slow one. So, in the same location where both fluids are found it
seems that the flow becomes multivalued~\citep{MHOC}. Nature resolves
this apparent contradiction by forming an initial discontinuity giving
rise to a hydrodynamical region separated by two shock waves $S$ and
a contact discontinuity $T$ between them~\citep{landaufm}. The contact
discontinuity $T$ is stationary with respect to the gas at both of its
sides. This set of discontinuities $S\_T\_S$ together with the gas that
is  between them is called a \emph{working surface} and is shown pictorially
in \figref{initial}.

\section{Relativistic strong shock waves}
\label{rel-shock}

  Consider an ejected one dimensional planar polytropic flow  with
a  pressure $p_{1}$, an internal energy density $\epsilon_{1}$ and a
mass density $\rho_{1}$. We assume the flow to have a time-variable 
ultra-relativistic
velocity $v_1 < 1$. This flow is injected into an external one that moves with
constant velocity $v_2<v_1$, a pressure $p_{2}$, an internal energy
density $\epsilon_{2}$ and mass density $\rho_{2}$. As explained in the
previous Section, the assumption of
supersonic ultra-relativistic periodic flow injected at the base of the
jet leads to the formation of a working surface since a given fast
flow parcel will eventually overtake a slow one, forming an initial 
discontinuity on the hydrodynamical variables.  All the discontinuities
(two shock waves and the contact discontinuity) and the flow between them,
inside the working surface move in the direction of the injected 
flow~\citep{landaufm,MHOC}.

\begin{figure}
\begin{center}
	\psfrag{v1}{$v_1$}
	\psfrag{vs}{$v_{s1}$}
	\psfrag{v3}{$v_{3}$}
	\psfrag{v4}{$v_{3'}$}
	\psfrag{y}{$\rho_{3'}, \epsilon_{3'}$}
	\psfrag{w}{$p_{3}=p_{3'}$}
	\psfrag{v2}{$v_2$}
	\psfrag{z}{$\rho_3, \epsilon_3$}
	\psfrag{vs2}{$v_{s2}$}
	\psfrag{x}{$p_{1}, \rho_{1}, \epsilon_{1}$}
	\psfrag{a}{$p_{2}, \rho_{2}, \epsilon_{2}$}
	\psfrag{S}{$S$}
	\psfrag{T}{$T$}
     \includegraphics[scale=0.50]{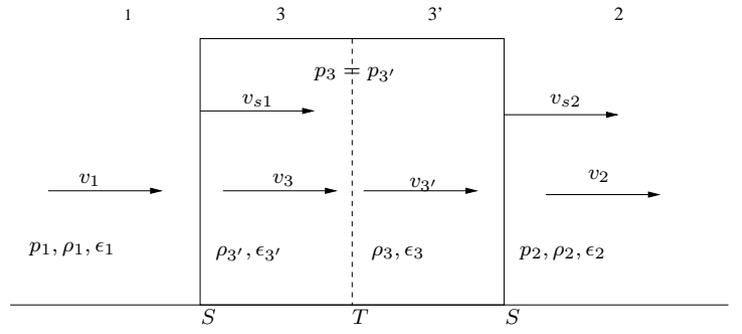}
\end{center}
	\caption[ Working surface within a relativistic flow.] { The
	figure shows a working surface, i.e. two shock waves separated
	by a contact discontinuity, which propagates into an ``external''
	moving gas with velocity $v_2$. It is assumed that the working surface
	has been generated by a flow with varying velocity \( v_1 \)
	injected from the
	left side of the figure. Solid vertical lines represent shock
	waves $S$ whereas the dashed one is the contact discontinuity
	$T$. The arrows indicate the direction of the flow velocity
	and the shock waves. The thermodynamical quantities $\rho$,
	$\epsilon$ and $p$ are different in each of the regions bounded
	by such discontinuities (except for the pressure \( p \) which
	is continuous across the contact discontinuity) and their values
	are determined from the initial conditions of the ejected flow
	together with the moving external initial flow. The velocity across
	the contact discontinuity is also constant so that \( v_3 = v_{3'}
	\), and is also the velocity of the contact discontinuity.}
\label{fig5}
\end{figure}

  \figref{fig5} shows a working surface, i.e.  two shock waves
separated by a contact discontinuity, such as one generated by
a supersonic periodic injection velocity. The external shock flow
between the contact discontinuity and the right hand flow is denoted
by region 3'. The shock wave at the right of the contact discontinuity
produces a discontinuity in the hydrodynamical quantities between
the external pre-shock flow ($p_{2}$, $\epsilon_{2}$, $\rho_{2}$,
$v_2$) and the post-shock one ($p_{3'}$, $\epsilon_{3'}$,  $\rho_{3'}$,
$v_{3'}$). Region 3 contains injected shock material with a mass density
different from the one in region 3'. Regions 3 and 3' are separated by
a tangential contact discontinuity.  The flow at the right and left of
the contact discontinuity and the contact discontinuity itself move at
the same velocity \( v_{3} = v_{3'} \) with a continuous pressure \(
p_3 = p_{3'} \).  However, the density is not continuous through the
contact surface and therefore neither the temperature, nor the entropy.
The  shock wave that separates region 1 from region 3 also produces a
strong discontinuity in the hydrodynamical quantities between the ejected
flow ($p_{1}$, $\epsilon_{1}$,  $\rho_{1}$, $v_1$) and the shock injected
gas ($p_{3}$, $\epsilon_{3}$ , $\rho_{3}$, $v_{3}$).

  The Taub shock wave jump conditions associated with mass, momentum
and energy for relativistic hydrodynamics~\citep[see
e.g.][]{taub48,taub78,landaufm} are calculated in a system of reference
for which a particular shock wave is at rest, and are given by:

\begin{align}
         v_\text{pre}\Gamma_\text{pre} \rho_\text{pre} &= v_\text{post}
	 \Gamma_\text{post} \rho_\text{post},\label{taub1}\\
	 \omega_\text{pre} v_\text{pre}^2 \Gamma_\text{pre}^2 + p_\text{pre} &=  \omega_\text{post} v_\text{post}^2 \Gamma_\text{post}^2 + p_\text{post}, \label{taub2}\\
	 \omega_\text{pre} v_\text{pre} \Gamma_\text{pre}^2 &= \omega_\text{post} v_\text{post} \Gamma_\text{post}^2.\label{taub3}
\end{align}

\noindent where $\Gamma_\text{pre,post} := 1/\sqrt{1 - v_\text{pre,post}^2
}$ are the Lorentz factors of the fluid velocities $v_\text{pre}$
and $v_\text{post}$ for any general pre-shock and
post-shock regions respectively.

  \citet{mccolg73} calculated general jump conditions valid for any
inertial system of reference, using conservation of relevant fluxes
through the shock wave's world-line in a four dimensional space-time.
These solutions seem to be adequate when dealing with
a system of reference in which the shock waves are moving, such
as the one shown in \figref{fig5}. However, in general terms
the resulting equations are lengthy and cumbersome to manipulate.
Particularly, for the problem we are interested in this article, the
manipulations become so complicated that we decided to proceed in a
different way. As described in the following sections we use  Taub's
jump conditions~\eqref{taub1}-\eqref{taub3} in the proper frame of
a particular shock wave and then perform the corresponding Lorentz
transformations to the final system of reference.

  To simplify the discussion and having in mind applications to high
energy phenomena in astrophysics, we assume now and in what follows that
both shock waves inside the working surface are strong.   To understand
what we mean by a strong shock in relativistic hydrodynamics let us
proceed in the following way, bearing in mind that strong means in
general terms that the post-shock pressure \( p_\text{post} \) is much
greater than the pre-shock one \( p_\text{pre} \), i.e. \( p_\text{post}
\gg p_\text{pre} \).  The pre-shock total energy density is given
by $e_{\text{pre}}= \rho_\text{pre} + \epsilon_\text{pre}$, where \(
\epsilon_\text{pre} \) is the pure internal thermal energy density of the
flow, and the total post-shock energy density is given by $e_\text{post}=
\rho_\text{post} + \epsilon_\text{post}$, with \( \epsilon_\text{post}
\) the internal thermal energy density of the corresponding flow.
In these terms, a strong shock must satisfy the following condition:
$\epsilon_\text{post} \gg \epsilon_\text{pre} $.  Furthermore, in the case
of ultra-relativistic fluids, the pre and post-shock rest energy density
is much smaller than each of their corresponding thermal energy densities,
i.e., $\epsilon_\text{pre} \gg \rho_\text{pro}$ and $\epsilon_\text{post}
\gg \rho_\text{post}$. This implies that the total energy densities
can be approximated as $e_\text{pre} \approx \epsilon_\text{pre}$,
$e_\text{post} \approx \epsilon_\text{post}$.

\section{Exact analytical solution.}
\label{analytic}

To find the exact solution of an ultra-relativistic working surface as
presented in Section~\ref{rel-shock} we proceed in the following way. Each of
the shock waves can be studied separately in their own system of reference
at rest using the Taub jump conditions \eqref{taub1}-\eqref{taub3}. The
jump in mass density trough the contact discontinuity merges naturally
through the union of both solutions after an appropriate set of Lorentz
transformation to the final system of reference.

In the proper system of reference of the shock waves, the pre-shock and
post-shock velocities are related to pure thermodynamical quantities
via the Landau-Lifshitz relations \citep{landaufm}, which for the case
we are studying can be written as:

\begin{equation}
  v_\text{prel}^2 = \frac{(p_3 - p_1)(e_3+p_1)}{(e_3 -e_1)(e_1 + p_3)}, \
    v_\text{prer}^2 = \frac{(p_{3'} - p_2)(e_{3'}+p_2)}{(e_{3'} -e_2)(e_2 +
    p_{3'})} ,
\end{equation}

\begin{equation}
  v_\text{postl}^2 = \frac{(p_3 - p_1)(e_1+p_3)}{(e_3 -e_ 1)(e_3 +
    p_1)}, \ v_\text{postr}^2 = \frac{(p_{3'} - p_2)(e_2+p_{3'})}{(e_{3'}
    -e_2)(e_{3'} + p_2)},
\end{equation}

\noindent where subindices 1 and 2 label pre-shock gas and subindices 3
and 3' represent the post-shock flow for the left (l subscript added)  
and right (r subscript added) shock wave
respectively according to \figref{fig5}.  Since we are interested
in ultra-relativistic flows with an ultra-relativistic equation of state,
then \( e = 3p \) and the previous set of equations approximated to \(
\mathcal{O}(e_\text{pre}/e_\text{post}) \) yield:

\begin{align}
  v_\text{pre} \approx  1 -
    \frac{4}{3}\frac{e_\text{pre}}{e_\text{post}}, \ \ 
  \Gamma_\text{pre}^2 = \frac{3}{8}\frac{e_\text{post}}{e_\text{pre}},
\label{G1}
\end{align}

\noindent and:

\begin{equation}
  v_\text{post} \approx \frac{1}{3}\left( 1+
    \frac{4}{3}\frac{e_\text{pre}}{e_\text{post}}\right),\ \
    \Gamma_\text{post}^2 \approx \frac{9}{8}.\label{G3}
\end{equation}

\begin{figure}
\begin{center}
	\psfrag{v1}{\tiny{\textbf{$v_1=v_{s1}\oplus v_\text{prel}$}}}
	%\psfrag{vs}{$v_{s1}$}
	\psfrag{v3}{\tiny{\textbf{$v_{3}=v_{s1}\oplus v_\text{postl}$}}}
	\psfrag{v4}{\tiny{\textbf{$v_{3'}=v_{s2} \ominus v_\text{postr}$}}}
	%\psfrag{y}{\small$\rho_3', e_3'$}
	\psfrag{vs} { $v_{s1}$}
	\psfrag{v2}{\tiny{\textbf{$v_2 = v_{s2} \ominus v_\text{prer}$}}}
	\psfrag{x}{$1$}
	\psfrag{a}{$2$}
	\psfrag{vs2}{$v_{s2}$}
	%\psfrag{2}{$Pre-choque$}
 \includegraphics[scale=0.50]{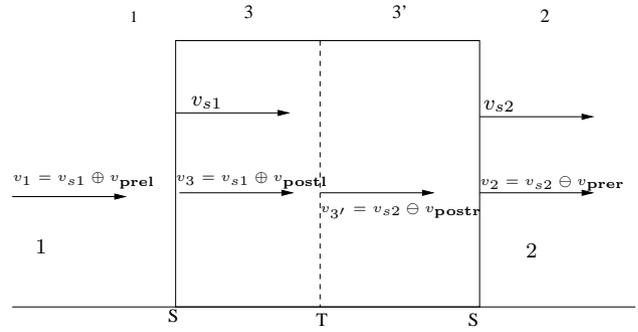}
\end{center}
	\caption[Change of reference system.]{By means of appropriate
	Lorentz transformations of velocities applied to the two (left and
	right) shock waves, we embed them into an adequate system of
	reference where the working surface moves to the right of the
	diagram.  The symbols $\oplus$
	and $\ominus$ represent the rule of addition and substraction 
	of relativistic velocities respectively.}
\label{sisref}
\end{figure}

  Let us now perform Lorentz transformations to each shock wave, so that
they can be accommodated into the appropriate description of a moving
working surface to the right.  To do so, let us add an ultra-relativistic
velocity $v_{s1} = 1 - \varepsilon_1$ to the left-hand shock wave where
$\varepsilon_1 \ll 1$.  We also add an ultra-relativistic velocity
$v_{s2}=1-\varepsilon_2$, with $\varepsilon_2 \ll 1$ to the right-hand
shock wave.  Note that the velocity \( v_{s2} \) has to be greater to
both  $v_\text{postr}$ and $v_\text{prer}$ in such a way that the flow
and the shock wave move to the right, consistent with the motion of the
working surface as a whole represented pictorially in  \figref{sisref}.

  In order to perform the required Lorentz transformations, we have
proved in the Appendix a set of very useful theorems relating the Lorentz
factor of relativistic additions valid for the case of ultra-relativistic
velocities.   In the following discussion we make extensive use of such
relations which greatly simplify our required approximations for the
Lorentz transformations as described in \figref{sisref}.

  To begin with, using equation~\eqref{a08} it follows that\footnote{In
what follows we use the addition \( \oplus \) or substraction \( \ominus \)
operators to mean relativistic addition or substraction of velocities}:

\begin{equation} 
\Gamma_{1}^2 = \Gamma_{s1 \oplus \text{prel}}^2 = 4 \Gamma_{s1}^2
\Gamma_\text{prel}^2,
\label{Gsuma1}
\end{equation}

\noindent where $\Gamma_i$ is the Lorentz factor of the velocity
$v_{i}$. Also, by means of relation~\eqref{v2little} applied to the
post-shock velocity \( v_\text{post} \) and the velocity \( v_\text{s1}
\) it follows that:

\begin{equation}
  \Gamma_{3}^2 = \Gamma_{s1 \oplus \text{postl}}^2= \frac{16}{9}
    \Gamma_{s1}^2 \Gamma_\text{postl}^2
    = 2 \Gamma_{s1}^2,
\label{Gs4} 
\end{equation}

\noindent according to the results of equation~\eqref{G3}.

  Now, the ratio of equation~\eqref{Gsuma1} to~\eqref{Gs4} together
with the substitution of \eqref{G1}, yields a jump condition relation
for the corresponding energy densities (or pressures) as follows:

\begin{equation}
  e_3 = \frac{4}{3}\frac{\Gamma_{1}^2}{\Gamma_{3}^2}e_{1}.
\label{p4}
\end{equation}

\noindent Note the necessary condition $\Gamma_{1} > \Gamma_{3}$
required in order for the entropy density to increase through the
shock wave.

  Finally, using equations \eqref{Gs4} and \eqref{G1} it is possible to
obtain a relation between the pre-shock  Lorentz factor $\Gamma_{1}$
and the one associated with the velocity of the left-hand shock wave
$\Gamma_{s1}$ given by:

\begin{equation}
  \Gamma_{s1}^2 = \frac{2}{3} \frac{e_{1}}{e_3}\Gamma_{1}^2.
\label{Gs1} 
\end{equation}

  We now turn our attention to the right-hand shock. Since $v_2 =
v_{s2} \ominus v_\text{prer}$, then $v_{s2} = v_2 \oplus v_\text{prer}$
and so:

\begin{equation}
  \Gamma_{s2}^2 = 4\Gamma_{2}^2 \Gamma_\text{prer}^2,
\label{Gsuma4} 
\end{equation}

\noindent according to equation \eqref{a08}.  Following the same procedure
as the one used to obtain relation~\eqref{Gs4} it is possible to show from
equation \eqref{v2little} that:

\begin{equation}
  \Gamma_{3'}^2  =  \frac{4}{9}\Gamma_{s2}^2 \Gamma_\text{postr}^2 =
    \frac{1}{2} \Gamma_{s2}^2,
\label{G4p} 
\end{equation}

\noindent The last step on the previous relation follows directly by
the application of equation~\eqref{G3}.

  In exactly the same form as the one for which  we obtained
equation~\eqref{p4}, substitution of equation~\eqref{G1} into the previous
relation yields:

\begin{equation}
  e_{3'} = \frac{4}{3}\frac{\Gamma_{3'}^2}{\Gamma_{2}^2}e_{2},
\label{p4p}
\end{equation}

\noindent and so, using relations~\eqref{G4p} and~\eqref{G1} it also
follows that:

\begin{equation}
  \Gamma_{s2}^2 = \frac{3}{2}\frac{e_{3'}}{e_{2}}\Gamma_{2}^2.
\label{Gs2}
\end{equation}

  To obtain a relation that describes the jump condition relating the
mass densities through a shock wave, we use the Taub adiabatic~\citep{taub48},
since it contains all thermodynamical information related to
pre-shock and post-shock quantities.  For the specific case of the
ultra-relativistic flow we are studying in this article, the
enthalpy per unit of volume is $w = 4p$, and so, the Taub adiabatic
becomes:

\begin{equation}
  4\frac{p_\text{pre}^2}{\rho_\text{pre}^2} -
4\frac{p_\text{post}^2}{\rho_\text{post}^2} + (p_\text{post}
- p_\text{pre})\left( \frac{p_\text{pre}}{\rho_\text{pre}^2} +
\frac{p_\text{post}}{\rho_\text{post}^2}\right)=0,
\label{ttaub2}
\end{equation}

\noindent i.e.:

\begin{equation}
  \frac{\rho_\text{post}}{\rho_\text{pre}}=
    \sqrt{\frac{p_\text{post}}{p_\text{pre}}\left(\frac{3 {p_\text{post}}
    / {p_\text{pre}}-1}{3+ {p_\text{post}}/{p_\text{pre}}} \right)}\approx
    \sqrt{3\frac{p_\text{post}}{p_\text{pre}}},
\label{Sdens}
\end{equation}

\noindent for a strong shock wave where $p_\text{post} \gg p_\text{pre}$.
Note that the ratio $ {\rho_\text{post}} / {\rho_\text{pre}} $
indicates that the post-shock mass density increases as the square root of
the compression factor \( p_\text{post} / p_\text{pre} \).  This is a
very well known fact of relativistic fluid dynamics and greatly differs
from the constant fixed value obtained for non-relativistic flows.

  Now, since the pressure across the contact discontinuity is continuous,
i.e. $p_{3}=p_{3'}$ -but not the mass density (and therefore neither
the temperature nor the entropy),  the movement of the flow is in
a single direction perpendicular to the contact discontinuity.  As a
result of this, the velocity is also continuous through this tangential
discontinuity, i.e.  $v_{3}=v_{3'}$ (and therefore its Lorentz factors
$\Gamma_{3} = \Gamma_{3'}$)~\citep{landaufm}.  Using all these facts, 
we can equate relations~\eqref{p4} and~\eqref{p4p} to obtain:

\begin{equation}
  \Gamma_{3}^2 = \Gamma_{3'}^2 = \Gamma_{1}\Gamma_{2}\sqrt{\frac{p_{1}}{p_{2}}},
\label{angamm}
\end{equation}

\noindent and so:

\begin{equation}
  v_3 = v_{3'} = \left(1 - \left(\frac{p_{2}}{p_{1}}\right)^{1/2}
    \left(1-v_1^2\right)^{1/2}\left(1-v_2^2\right)^{1/2}\right)^{1/2}.
\label{anvel}
\end{equation}

\noindent Since a necessary condition for the working surface to be 
formed requires
\( p_1 > p_2 \) then \( \Gamma_3^2 > \Gamma_1 \Gamma_2 \) and since the boundary values
for the Lorentz factors satisfy \( \Gamma_1 >1 \) and \( \Gamma_2 >1 \) then \( \Gamma_3 > 1 \) 
as expected.

\noindent Also, substitution of equation~\eqref{angamm} into~\eqref{p4p}
yields:

\begin{equation}
  p_3 = p_{3'} = \frac{4}{3}\frac{\Gamma_{1}}{\Gamma_{2}}\left(
    p_{1}p_{2}\right)^{1/2}.
\label{anp}
\end{equation}

Finally, using equations \eqref{Sdens} and \eqref{anp}, the jump in the mass 
density through the contact discontinuity is given by:

\begin{equation}
    \frac{\rho_3}{\rho_1} = \sqrt{4\frac{\Gamma_1}{\Gamma_2}} \left(\frac{p_2}{p_1} \right)^{1/4},
\label{Sdens1}
\end{equation}

\noindent and:

\begin{equation}
    \frac{\rho_3'}{\rho_2} = \sqrt{4\frac{\Gamma_1}{\Gamma_2}} \left(\frac{p_2}{p_1} \right)^{1/4}.
\label{Sdens2}
\end{equation}

  For completeness, using equations~\eqref{Gs1} and~\eqref{Gs2}, we 
express the Lorentz factors of the shock waves
as a function of the external hydrodynamical quantities:

\begin{equation}
  \Gamma_{s1}^2 = \frac{ 1 }{ 2 } \Gamma_1 \Gamma_2 \sqrt{p_1/p_2},
\label{Gs1fin}  
\end{equation}

\begin{equation}
  \Gamma_{s2}^2 = 2 \Gamma_1 \Gamma_2 \sqrt{p_1/p_2},
\label{Gs2fin}  
\end{equation}

\noindent which imply:

\begin{equation}
  \Gamma_{s2} = 2 \Gamma_{s1}.
\label{Gsrelations} 
\end{equation}

\noindent  The corresponding shock wave velocities are then given by:

\begin{equation}
  v_{s1} = 1 - \frac{ 1  }{ \Gamma_1 \Gamma_2 } \sqrt{ \frac{ p_2 }{ p_1 } },
\label{s1vel} 
\end{equation}

\begin{equation}
  v_{s2} = 1 - \frac{ 1  }{ 4 \Gamma_1 \Gamma_2 } \sqrt{ \frac{ p_2 }{ p_1 } }.
\label{s2vel} 
\end{equation}

\noindent The ratio between these last two equations yield 

\begin{equation}
  v_{s2} =  \frac{ v_{s1} }{ 4 } + \frac{ 3 }{ 4 },
\label{shockspeeds} 
\end{equation}

\noindent and since \( v_{s1} < 3/4 + v_{s1}/4 \)  for \( v_{s1} < 1 \), it
then follows  that \( v_{s2} > v_{s1} \) as expected.

  Equations~\eqref{angamm}-\eqref{Gs2fin} are the
required solution we are looking for.  They are a set of relationships
that determine the complete analytical solution of an ultra-relativistic
planar working surface in one dimension moving to the right.  The jumps
in the hydrodynamical quantities (pressure --or total energy density, 
mass density and velocity --or Lorentz factor) are completely determined
by the boundary conditions, i.e. the external hydrodynamical quantities
to the working surface (labelled 1 and 2).

\section{Hydrodynamical profiles}
\label{profiles}

  We have implemented a numerical code with the obtained hydrodynamic
conditions \eqref{Sdens}, \eqref{anvel} and \eqref{anp} in such a way
that the working surface evolves in position and time.  The positions
\(x \) of the shock waves evolve strictly in time \( t \) with the
velocities of the shock waves given by~\eqref{Gs1} and~\eqref{Gs2} following
the characteristic lines $\mathrm{d}x/ \mathrm{d}t = ( c_\text{s} \pm
v) /(1 \pm c_\text{s}v)$ \citep[see e.g.][]{mendoza00}, where $
c_\text{s}=1/\sqrt{3}$ is the speed of sound for an ultra-relativistic
equation of state.  The contact discontinuity travels with the velocity
of the flow around it, i.e.  with velocity~\eqref{anvel} and so, its
position at time \( t \) moves as \( v_3 \, t \) from the starting point.

\begin{figure}
      \begin{center}
    \includegraphics[angle=0, width=8.6cm]{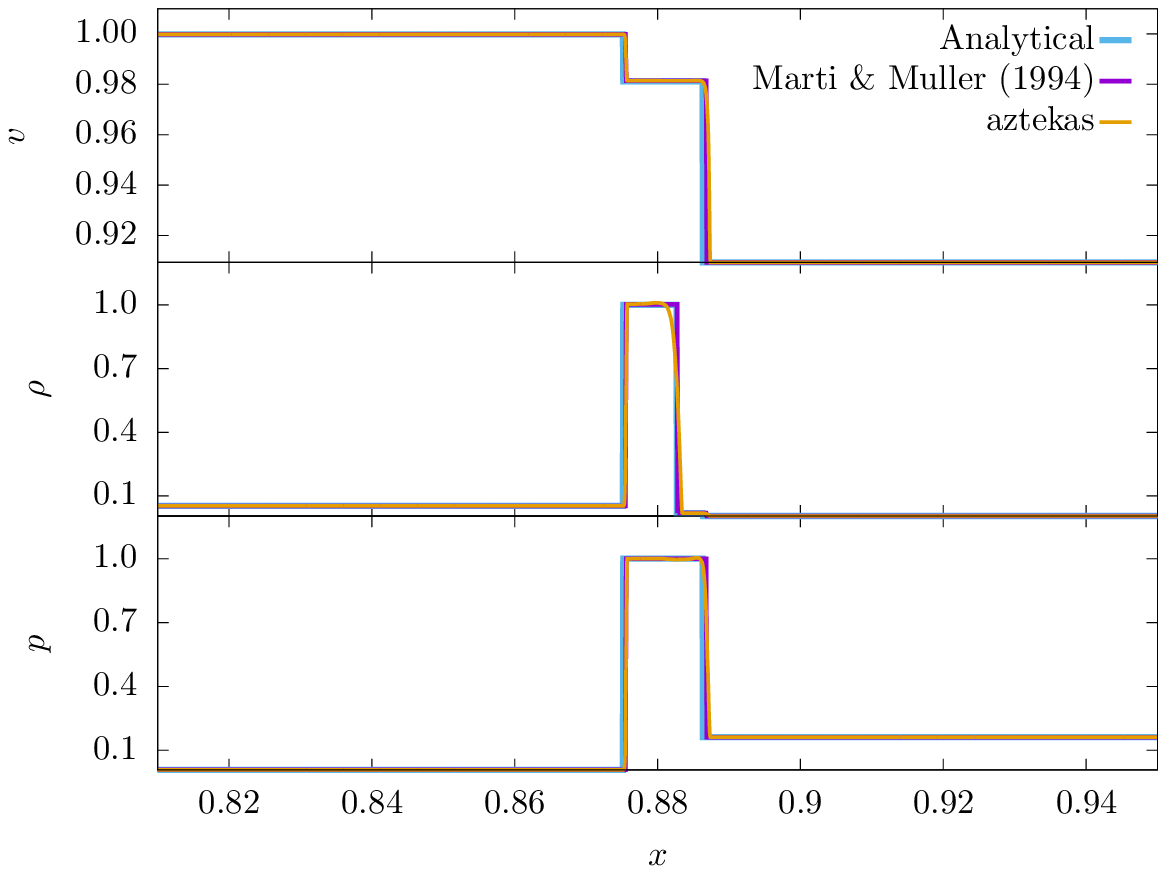}
	\includegraphics[angle=0, width=8.6cm]{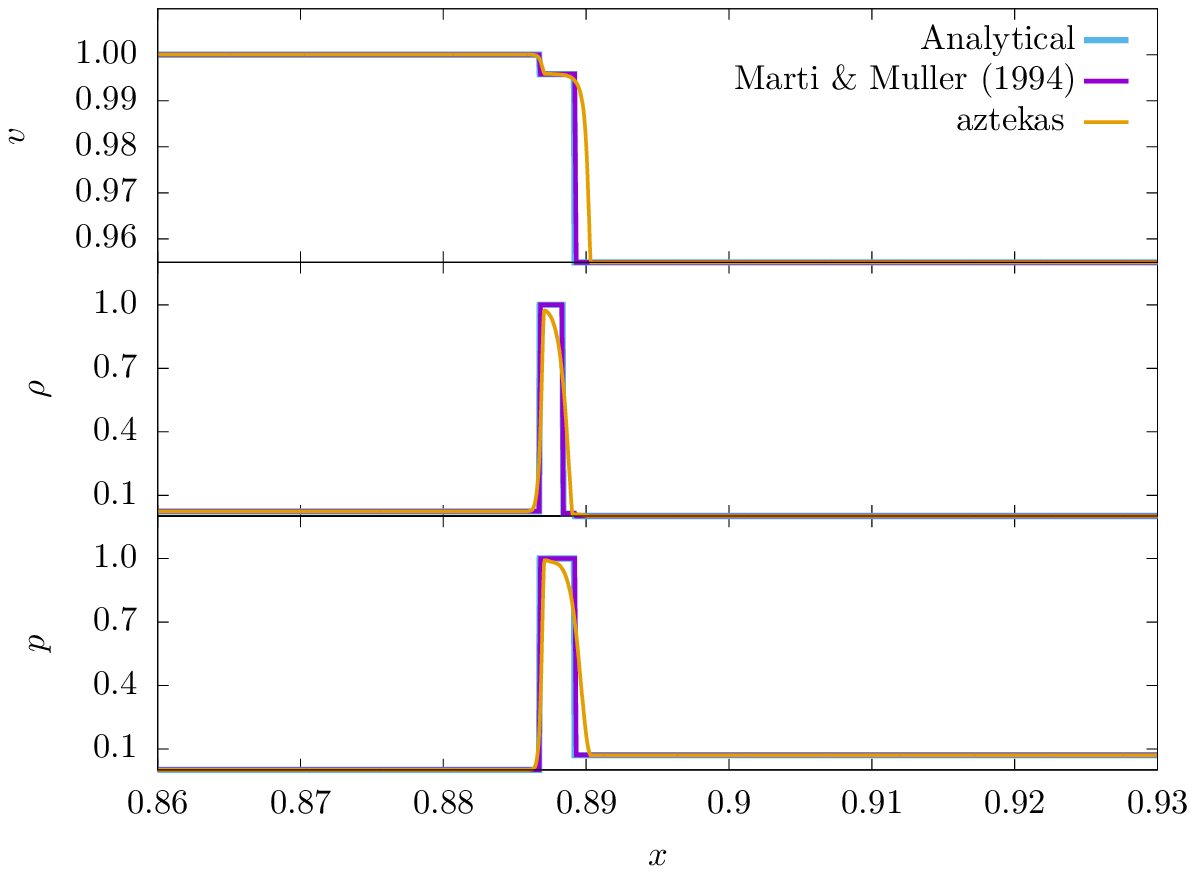}\\
	\includegraphics[angle=0, width=8.6cm]{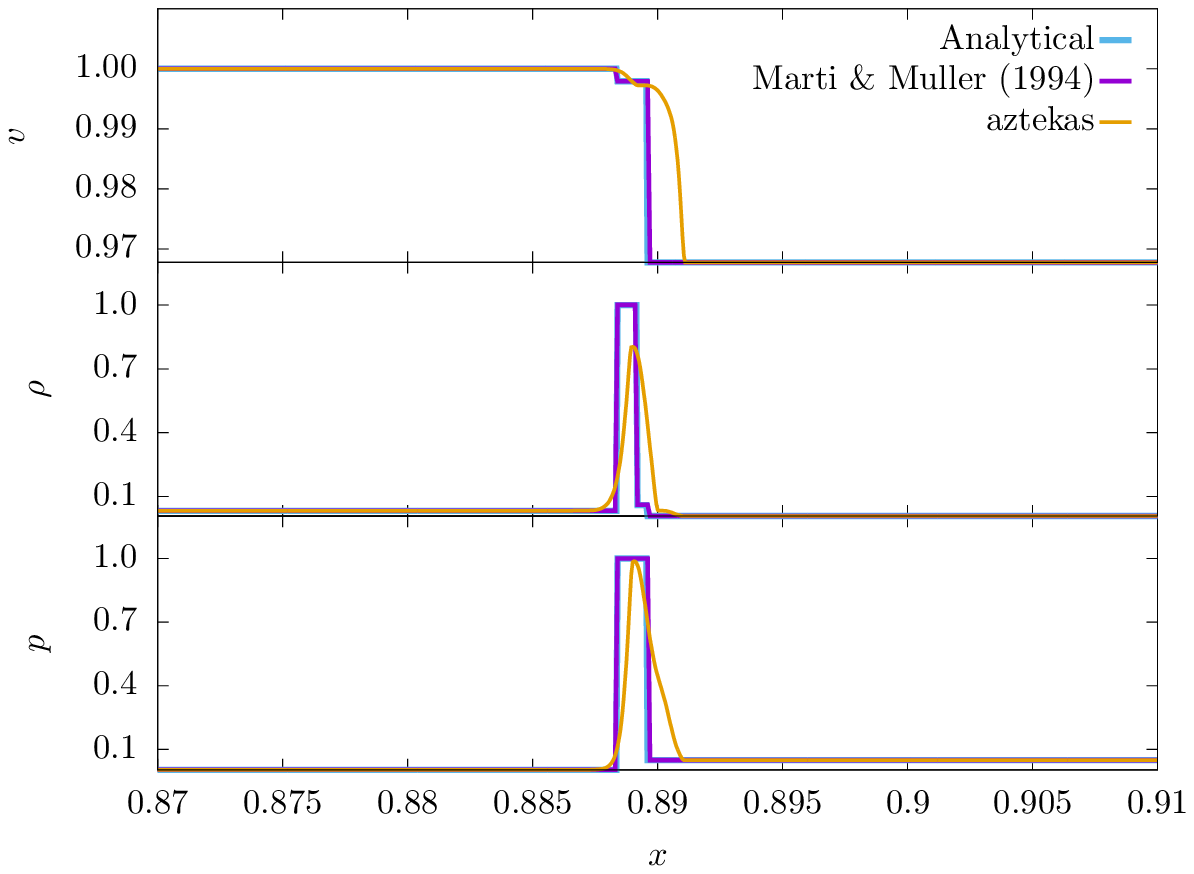}\\
       \end{center}
	\caption[Hydrodynamical profiles.]{ 
	 Snapshot at time \( t = 0.39 \)  for the 
	 hydrodynamical quantities of a working surface that moves
	 to the right. 	 From top to bottom the plots present pressure \( p
	 \), mass density \( \rho \) and velocity \( v \) profiles as a
	 function of position.
	 Continuous blue and purple lines  represent the analytical (this
	 article) and semi-analytic~\citep{marti94} solutions respectively (both merge
	 together).  Orange lines represent a 1DRHD numerical simulation
	 using the aztekas code (\url{http://aztekas.org} \copyright 2008
	 Sergio Mendoza \& Daniel Olvera and \copyright 2018 Alejandro
	 Aguayo-Ortiz \& Sergio Mendoza).  
	 Horizontal lines represent the spatial position of the two shock
	 waves. The horizontal axis scale was chosen in such a way 
	 that the initial discontinuity at time \( t= 0\) occured at \( x = 0.5 \) on a 
	 ``numerical box'' for which  \( 0 \leq x \leq 1 \). The vertical axis was 
	 chosen in such a way that the particular hydrodynamical quantity is normalised 
	 to its maximum value for each particular panel.  The analytic solution presented in this
	 article and the semi-analytic one of \citet{marti94} are almost identical to one another.}
\label{fig7}
\end{figure}

 \figref{fig7} shows pressure, mass density and velocity profiles at
different times for different initial hydrodynamical parameters. The plots
were normalised in such a way that the horizontal axis varies from  \(
0 \) to \( 1 \) and the vertical axis values are normalised to \( 1 \)
with respect to the maximum value of the corresponding hydrodynamical
quantity.  All plots are fixed snap shots at time $t=0.39$ after the
formation of the initial discontinuity which is assumed to occur at time
\( t = 0 \)  when the flow was initiallly discontinuous at a 
certain position with left and right fixed hydrodynamical values given in \autoref{table2}.  

\begin{table}
  \begin{tabular}{|c|c|c|c|}
    \hline 
    Case & \( p_\text{left} \) & \( p_\text{right}\) & \( \rho_\text{left} \) \\ 
    \hline
    (a) & \( 1 \times 10^3 \) & \( 2 \times 10^4 \) & \( 0.1 \) \\
    (b) & \( 1 \times 10^3 \) & \( 4 \times 10^4 \) & \( 0.1 \)  \\
    (c) & \( 2 \times 10^3 \) & \( 3 \times 10^4 \) & \( 0.8 \) \\
    \hline
     Case & \( \rho_\text{right}\)  & \(v_\text{left} \ (\Gamma_\text{left}) \) & \( v_\text{right} \ (\Gamma_\text{right})\) \\
     \hline
     (a) & \( 0.01 \) & \(0.999799  \ (\approx 50) \) & \( 0.909355 \ (\approx 2.4) \) \\
     (b) & \( 0.01\) & \( 0.99999 \ (\approx 223.00)\) & \(0.717315 \ (\approx 1.43) \) \\
     (c) & \( 0.2 \) & \( 0.999991 \ (\approx 235.7)\) & \( 0.96778 \ (\approx 4.0) \) \\
    \hline
\end{tabular}
\caption[Initial conditions.]{The table shows three different cases
with initial conditions in the hydrodynamical quantities $p$, $\rho$, $v$ and $\Gamma$
associated with the right and left pre-shock flows that form the working surface represented 
by the analytical and numerical simulations described in \autoref{fig7}.}
\label{table2}
\end{table}

The figure shows three models: (1) our full analytical model
discussed in this article (continuous line), (2) the results of an 1DRHD
numerical simulation using  the Free GNU Public Licensed code aztekas
(\url{https://aztekas.org} \copyright 2008 Sergio Mendoza \& Daniel Olvera
and \copyright 2018 Alejandro Aguayo-Ortiz \& Sergio Mendoza).  This code
uses the Primitive Variable Recovering Scheme (PVRS) to obtain directly
the primitive variables of the hydrodynamical problem~\citep{aguayo2018}
and (3) the semi-analytical solution of~\citet{marti94}.

  From the results of \figref{fig7} it can be seen that the
semi-analytical solution of \citet{marti94} and our full exact solution
are in complete agreement, which can also be verified by the results
of \autoref{table1}, which shows the relative error between both
solutions.  Note that the numerical solution fails to reach the analytic
and semi-analytical profiles at large Lorentz factors.
This is a common and unsolved problem to all current numerical codes.

\begin{table}
  \begin{tabular}{|c|c|c|c|c|}
    \hline 
    Case &  \( \delta \rho \)	&  \( \delta p \) & \( \delta v \)  & \(  \delta (\Gamma v) \) \\
    \hline 
    (a)	& \( \lesssim 2.7 \times 10^{-3} \) & \( \lesssim 1.6
      \times 10^{-5} \) & \( \lesssim 2.6 \times 10^{-4} \) &  \( \lesssim 6.9 \times 10^{-3} \)  \\
    (b)	& \( \lesssim 5.8 \times 10^{-4} \) & \( \lesssim 6.1
      \times 10^{-7} \) & \( \lesssim 6.8 \times 10^{-6} \)  &  \( \lesssim 8.1 \times 10^{-4} \) \\
    (c)	& \( \lesssim 1.0 \times 10^{-3} \) & \( \lesssim 6.5
      \times 10^{-7} \) & \( \lesssim 2.0 \times 10^{-8} \) & \( \lesssim 4.8 \times 10^{-6} \) \\
    \hline
  \end{tabular}
  \caption[Comparison with semi-analytical model]{ The Table shows the
	  maximum relative error 
	  \( \delta X := | X_\text{M} - X | / X_\text{M}
	  \), where X stands for mass density \( \rho \), pressure \(
	  p \), velocity \( v \) and the product \( \Gamma v \). 
	  The subindex M refers to the
	  semi-analytical solution of~\citet{marti94}. Quantities without
	  subindex are our full ultra-relativistic analytical solution. Cases
	  (a), (b) and (c) refer to the numerical simulations performed
	  in \figref{fig7}. We also compared our results with much
	  greater initial pressures, densities and velocities as the
	  ones corresponding to those cases, reaching Lorentz factors \(
	  \sim 700 \) obtaining \( \delta \rho \lesssim 3.6 \times
	  10^{-4} \), \( \delta p \lesssim 4.6 \times 10^{-7} \) and \(
	  \delta v \lesssim 8.1 \times 10^{-6} \).
          }
\label{table1}
\end{table}

\section{Energy inside the working surface }
\label{workingsurface-energy}

  The total energy \( E(t) \) inside the working surface, i.e. between both shocks, is an 
important quantity to take into account for astrophysical processes, since the gas
inside it has been heated through two strong shock waves and it is expected to cool
via some efficient radiation process.  To take into account this full radiation process is outside
the scope of the present article.  However, we can make a few assumptions in order to have 
a first glance as to the relevance of such an important process that occurs inside relativistic
astrophysical jets.  In what follows we assume that the total energy $E(t)$ is radiated away
completely from the working surface due to some very efficient cooling
process and so, it is possible to calculate the energy loss or the
luminosity \( L(t) \)  by  computing the negative 
power $-\mathrm{d}E/\mathrm{d}t$ emitted by the working surface.  
This assumption is a first approximation to the problem since a fluid particle crossing
any of the shock waves and radiating all its energy ceases to have physical meaning, and 
will be sufficient for the first glance approximation we are dealing with.  We also assume 
that at a certain time \( t \) the energy density between the left shock wave \( S_1 \) and 
the tangential discontinuity \( T \) is only given by \( e_3(t) \), and that between the 
tangential discontinuity \( T \) and the right shock wave \( S_1 \) is given by \( e_{3'}(t) \). 
This is a simple approximation since formally both energy densities \( e_3 \) and \( e_{3'}\) are
functions of the position \( z \) along the injected direction of the flow and time \( t \).  
Nevertheless, since our first assumption is that the total energy within the working surface is
fully radiated away, this statement constitutes a first good approximation.

\subsection{Periodic velocity.}
\label{examples}

  In order to show how to compute the emitted power by the working
surface for a simple model, let us assume that the flow injection 
is periodic following the
assumptions of~\citet{MHOC}, with a periodic injected velocity given by:

\begin{equation}
  v_1=v_{10}+v_{11}\sin(\omega t),
\label{eq10.7}
\end{equation}

\noindent with $v_2=\text{const.}$, together with a constant
background injected velocity  $v_{10}$ close to the speed of light.
The small positive parameter $v_{11} \ll 1$ and $\omega$ is the angular
frequency of the oscillating flow. This choice produces a small sinusoidal
perturbation about a background velocity $v_{10}$ close to the speed
of light. The parameter $v_{11}\sin(\omega t)$ is chosen small enough
so that the motion of the flow is always subluminal.  Note that since the maximum value 
for the velocity is obtained when \( \sin(\omega t = 1) \), then a subluminal
velocity \( v_1 \) at that maximum value is such that \( 1 - v_1 |_\text{max} = 1 - v_{10}
- v_{11} > 0 \), so that \( v_{11} / \left( 1 - v_{10} \right) < 1  \).  In other words,
the Lorentz factor associated to the velocity \( v_1 \) is given by:

\begin{equation}
  \Gamma_1 = \frac{ 1 }{ \sqrt{ 2 \left( 1 - v_{10} \right) } } \left\{ 1 - 
    \frac{ v_{11} }{ 2 \left( 1 - v_{10} \right) } \sin( \omega t ) \right\}.
\label{G1sin}
\end{equation}

The total energy \( E \) inside the 
working surface, i.e. between both shock waves, is given by the volume \( V \) integral:

\begin{equation}
  E = \int_{V_1} e_3 \, \mathrm{d}V + \int_{V_2} e_{3'} \, \mathrm{d}V.
\label{energy}    
\end{equation}

\noindent In the previous equation, the volume \( V_1 \) represents the volume between the 
left shock wave \( S_1 \) and the contact discontinuity \( T \) with a transverse 
area \( A \).  The 
volume \( V_2 \) is taken from the tangential discontinuity \( T \) to the right shock wave 
\( S_2 \)  with the same transverse area \( A \).   Since the flow has plannar symmetry, 
equation~\eqref{energy} can be simplified as:

\begin{equation}
  E = A \int_{S_1}^{T} e_3 \, \mathrm{d}z + A \int_{T}^{S_2} e_{3'} \, \mathrm{d}V = 
      A \int_{S_1}^{S_2} e_3 \mathrm{d} z.
\label{energy2}    
\end{equation}

\begin{figure}
  \begin{center}
    \includegraphics[scale=0.56]{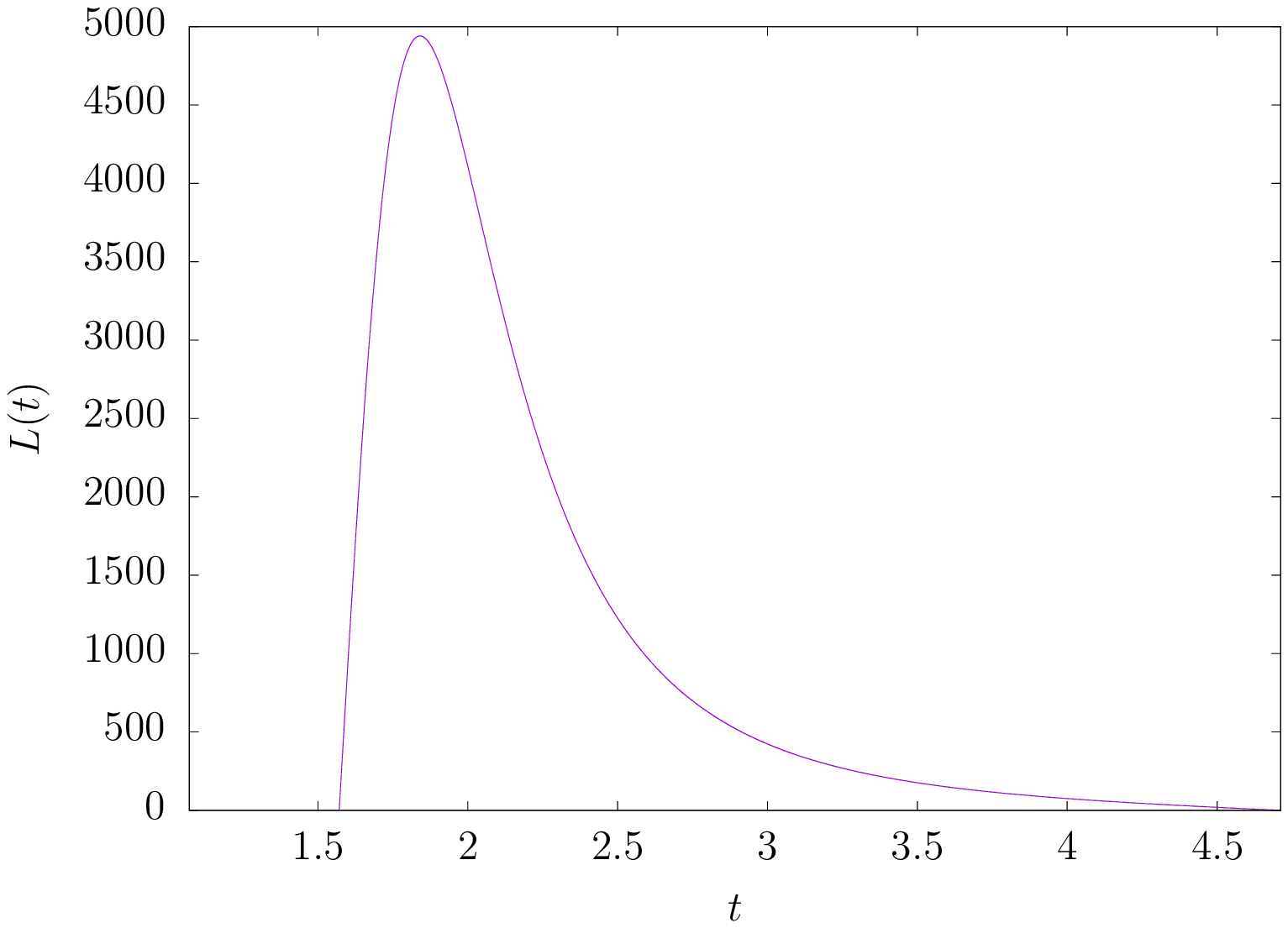}
    \includegraphics[scale=0.56]{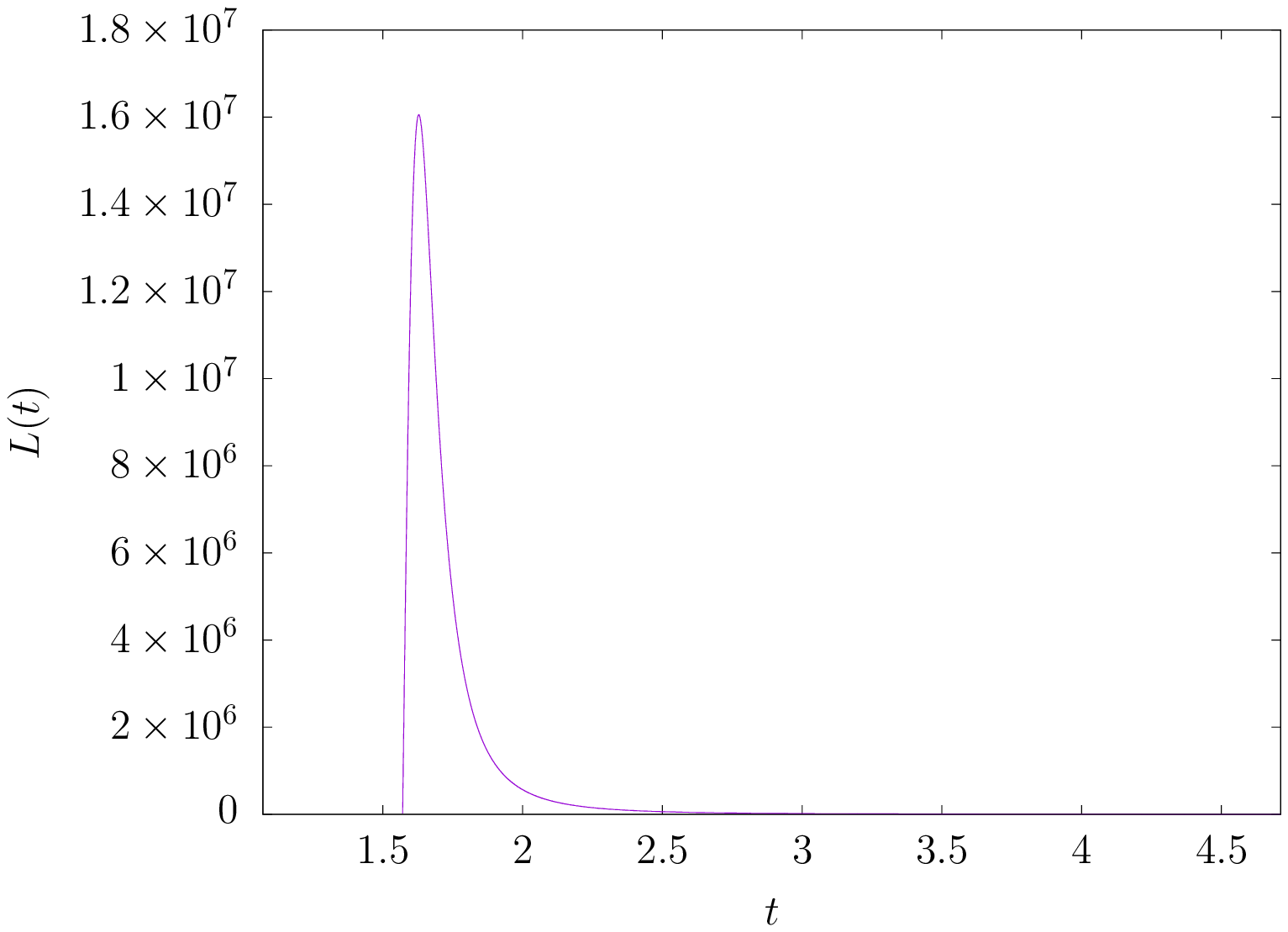}
    \includegraphics[scale=0.56]{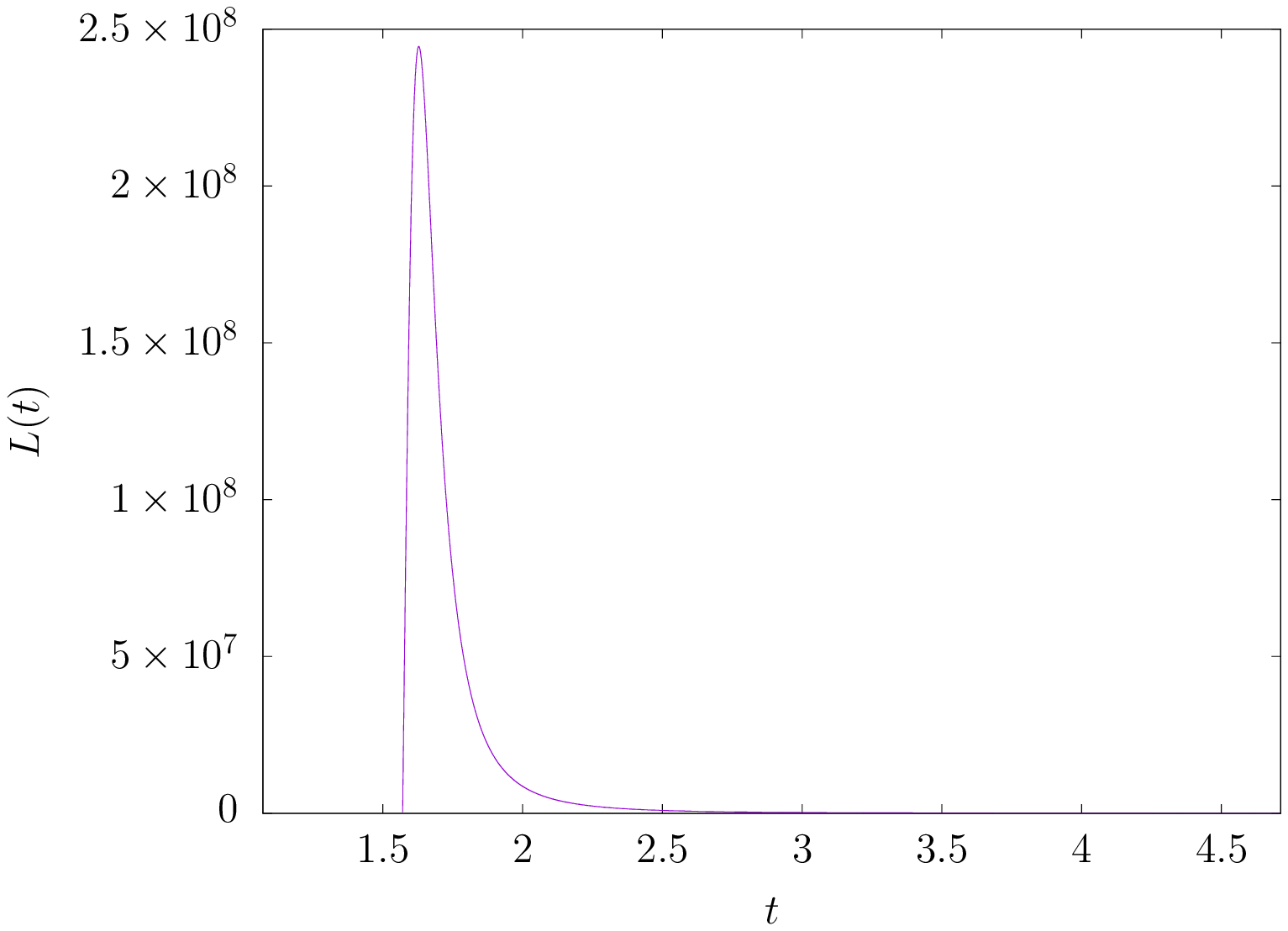}
  \end{center}
  \caption[Light curves]{ The figure shows three plots of the dissipated
	      energy density inside a working surface
	      %(assuming a perfectly efficient cooling mechanism).
	      From top to bottom the plots were produced with the following initial
	      conditions: (1) $v_{10}=0.999$, $v_{11}=0.0009$, 
	      $p_{1}=30000.0$, $p_{2} = 1000.0$, $v_2=0.1$; 
	      (2) $v_{10}=0.999998$, $v_{11}=0.00000199$, 
	      $p_{1}=40000.0$, $p_{2} = 1000.0$, $v_2=0.96$; 
	      (3) $v_{10}=0.9999998$, $v_{11}=0.000000199$, 
	      $p_{1}=100000.0$, $p_{2} = 90000.0$, $v_2=0.99$. In all
	      plots we set \( \omega = 1 \), \( \Delta z_\text{init} = 1 \) 
	      and chose a system of units 
	      where the velocity is measured in units of the speed of light.   The luminosity 
	      function \( L(t) \) has been normalised to the background injected mechanical 
	      power \( A p_1 v_{10} \),  for a transverse area \( A = 1 \),
	      and to the efficiency  conversion factor \( \eta \).
          }
\label{fig7.1}
\end{figure}

\noindent 
The last step in the previous equation follows from the fact that \( e_3 = e_{3'} \).
In what follows and for the purpose of the current calculation, we set the transverse area 
\( A = 1 \) and since in our model \( e_3 = e_3(t) \) only as stated above, 
the previous equation means that:

\begin{equation}
  E = e_3 \, \Delta z,
\label{energy-fin}
\end{equation}

\noindent  where \( \Delta z \) is the 
distance between both shock waves.  
To compute this length, we substitute equation~\eqref{G1sin} into equations~\eqref{s1vel} 
and \eqref{s2vel} to obtain a differential equation for the positions \( z_1 \) and \( z_2 \) 
of the left and right shock waves respectively:

\begin{gather}
  \frac{ \mathrm{d} z_1 }{ \mathrm{d} t} = v_{s1} = 1 - \frac{ \sqrt{ 2 \left( 1 - v_{10} 
    \right) }  }{ \Gamma_2 } \sqrt{ \frac{ p_2 }{ p_1 } }  \left\{ 1 + \frac{ v_{11} 
    \sin( \omega t ) }{ 2 \left( 1 - v_{10} \right) } \right\}, 
                         \label{vs1sin} \\
  \frac{ \mathrm{d} z_2 }{ \mathrm{d} t} = v_{s1} = 1 - \frac{ \sqrt{ 2 \left( 1 - v_{10} 
    \right) }  }{ 4 \Gamma_2 } \sqrt{ \frac{ p_2 }{ p_1 } }  \left\{ 1 + \frac{ v_{11} 
    \sin( \omega t ) }{ 2 \left( 1 - v_{10} \right) } \right\}, 
                         \label{vs2sin} 
\end{gather}

\noindent so that:

\begin{gather}
  z_1 = t - \frac{ \sqrt{ 2 \left( 1 - v_{10} 
    \right) }  }{ \Gamma_2 } \sqrt{ \frac{ p_2 }{ p_1 } }  \left\{ t - \frac{ v_{11} 
    \cos( \omega t ) }{ 2 \omega \left( 1 - v_{10} \right) } \right\} + z_{1\text{init}}, 
                         \label{z1} \\
  z_2 = t - \frac{ \sqrt{ 2 \left( 1 - v_{10} 
    \right) }  }{ 4 \Gamma_2 } \sqrt{ \frac{ p_2 }{ p_1 } }  \left\{ t + \frac{ v_{11} 
    \cos( \omega t ) }{ 2 \omega \left( 1 - v_{10} \right) } \right\} + z_{2\text{init}}, 
                         \label{z2} 
\end{gather}

\noindent where \( z_{1\text{init}}\) and \( z_{2\text{init}} \) are 
the initial poistions of the left and right shock waves respectively. With all the above 
relations and assuming a constant mass density and pressure discharges, the power loss inside 
the working surface is then given by:

\begin{equation}
\begin{split}
  L(t) &= - \eta \, \frac{ \mathrm{d} E }{ \mathrm{d} t } = 
   \eta \, \frac{ \mathrm{d} \Delta z }{ 
    \mathrm{d} t } \, e_3 +  \eta \, \frac{ \mathrm{d} e_3 }{\mathrm{d} t } \, \Delta z, 
                \\
    &= - 3 \eta \sqrt{2} \, P_2 \sqrt{1 - v_{10} } \left\{ 1 + \frac{ v_{11} \sin( \omega t ) 
      }{ 2 \left( 1 - v_{10} \right) } \right\} -  8 \eta \omega v_{11} \, \times 
                \\
    &  \left( \frac{ \Gamma_1^2(t)}{ \Gamma_2 } \right)^2 \sqrt{ p_1 p_2 } 
      \left\{ \frac{ 3 \left( 1 - v_{10} \right) }{ 
      2 \sqrt{2} \, \Gamma_2 } \sqrt{\frac{p_2}{p_1}
      } \left( t - \frac{ \cos( \omega t ) }{ 2 \omega \left( 1 - v_{10} \right) } \right) + 
      \right.
                \\
    & \Delta z_\text{init} \bigg\},
\end{split}    
\label{luminosity}
\end{equation}

\noindent where \( \Delta z_\text{init} \) 
is the initial length separation of both 
shock waves exactly at the time the working surface is formed and \( \eta \) is
an efficiency conversion factor from thermal energy power \( - \mathrm{d}{E}/\mathrm{d}{t} 
\) to luminosity \( L(t) \), such that \( 0 < \eta \leq 1 \). 
\figref{fig7.1} shows different power light curves obtained using
this last result for various hydrodynamical  parameters, chosen 
in such a way as to see the relevance of a high initial discontinuity between 
the pressures \( p_1 \)  and \( p_2 \).

\section{Discussion}
\label{discussion}

  In this work we have found ultra-relativistic analytical solutions for a
1D planar working surface (two shock waves separating from a contact
discontinuity) expanding into a moving medium.  The equation of
state of the flows outside and inside the working surface was also assumed to
be ultra-relativistic, i.e. with the internal energy density being much
greater than the rest mass internal energy density, and with a polytropic
index \( \kappa = 4/3 \) valid for ultra-relativistic gases.  In general
terms, this solution constitutes an analytical solution of the
ultra-relativistic shock tube problem when e.g. fast flow reaches 
slower one in a 1D planar setup.

This type of solution can be well adapted to the blobs and knots observed
in relativistic jets in Active Galactic Nuclei (AGN), microquasars
and long Gamma-Ray Bursts (lGRB).  In other words, the knots or blobs
observed in many of these astrophysical jets can be interpreted as
working surfaces moving along the jet.  In order to calculate the energy
dissipated inside the working surface as a function of time, we computed
the energy density loss (luminosity density) as a function of time for a
very particular model in which an ultra-relativistic varying velocity flow
with constant discharge is assumed to be injected at the base of the jet.
The luminosity density profiles produced by this efficient mechanism have
similar shapes to the ones observed on the light curves of outbursts in
AGN, microquasars and the prompt emission of lGBR.  It is our intention
in future works to use this model in order to understand and interpret
in detail the light curves of many of these astrophysical jets using
the methods of \citet{MHOC,Benitez2013,coronado2015,coronado2016},
but now with the full hydrodynamical solution presented in this article.

\section*{Acknowledgements}
\label{acknowledgements}
We thank an anonymous referee for the very useful comments 
made to the manuscript.
This work was supported by a PAPIIT DGAPA-UNAM grant IN112019 and
a CONACyT grant (CB-2014-1 No. 240512).  SM acknowledge support from
CONACyT (26344).

%%%%%%%%%%%%%%%%%%%%%%%%%%%%%%%%%%%%% 

\section*{Data Availability}
The data underlying this article are available in the article and in
its online supplementary material.



\bibliographystyle{mnras}
\bibliography{delacruz-mendoza}

%%%%%%%%%%%%%%%%%%%%%%%%%%%%%%%%%%%%%%%%%%%%%%%%%%

%%%%%%%%%%%%%%%%% APPENDICES %%%%%%%%%%%%%%%%%%%%%

\appendix 

\section{Lorentz factor relations for ultra-relativistic velocities}
\label{appendix01}

  In this section, we show how to obtain a set of useful formulae in which
Lorentz factors in different frames of reference are shown to be related to
one another for the cases in which the velocities they refer to are
highly relativistic.

\begin{theorem}
  If  $v_1$ and $v_2$ are two completely arbitrary velocities in units of
the speed of light that satisfy the rule of velocity addition

\begin{equation}
  v_2 \opm v_1 = \frac{v_2 \pm v_1}{1 \pm v_2v_1},
\label{apvel1}
\end{equation}

\noindent then, the Lorentz factor associated with this transformation is:

\begin{equation}
  \Gamma^2_{v_2 \opm v_1} =  \Gamma_{1}^2\Gamma_{2}^2(1 \pm v_2 v_1)^2.
\label{apgamm1}
\end{equation}

\end{theorem}

\noindent To demonstrate this property note that:

\begin{equation*}
 \begin{split}
  \Gamma^2_{v_2 \opm v_1} 
    &= \frac{1}{1 - \left(\frac{v_2 \pm v_1}{1 \pm v_2v_1}\right)^2} =
      \frac{(1 \pm v_2v_1)^2}{(1 \pm v_2v_1)^2- (v_2 \pm v_1)^2},  \\
    &= \frac{(1 \pm v_2v_1)^2}{1 + v_2^2v_1^2-v_2^2-v_1^2} = \frac{(1
	\pm v_2v_1)^2}{(1-v_1^2)(1-v_2^2)}.
\end{split}
\end{equation*}

  Using relation \eqref{apgamm1}, the following important properties of
the Lorentz factors can be obtained:

\begin{theorem}
\label{teoremita}
If  $\Gamma_{1}\gg1$ and $\Gamma_{2}\gg1$, that is, $v_{1,2} =
1- \varepsilon_{1,2}$, with  $\varepsilon_{1,2}\ll1$ sufficiently
small quantities that make the velocities $v_1$ and  $v_2$ to be
ultra-relativistic, then

\begin{equation}
  \Gamma_{1,2} = \frac{1}{ \sqrt{ 2\, \varepsilon_{1,2}}},
\label{tt}
\end{equation}

\noindent which combined with relation \eqref{apgamm1} yields:

\begin{align}
  \Gamma_{v_2 \oplus v_1} 
    &=2\Gamma_1\Gamma_2
      -\frac{1}{2}\left(\frac{\Gamma_{2}}{\Gamma_{1}}+
      \frac{\Gamma_{1}}{\Gamma_2}\right), \nonumber \\
  \Gamma_{v_2 \ominus v_1} &=
    \frac{1}{2}\left(\frac{\Gamma_{2}}{\Gamma_{1}}+
    \frac{\Gamma_{1}}{\Gamma_2}\right).
\label{prop1} 
\end{align}

\end{theorem}

 To proove this Theorem note first that equation~\eqref{tt} follows
directly from the required assumptions. The set~\eqref{prop1} can be
obtained from taking the square root of equation~\eqref{apgamm1}:

\begin{equation*}
  \begin{split}
	\Gamma_{v_2 \opm v_1} &= \Gamma_1\Gamma_2\left(1\pm(1-\varepsilon_1)(1-\varepsilon_2) \right),  \\
	&=\Gamma_1\Gamma_2\left( 1\pm 1\mp (\varepsilon_1 + \varepsilon_2) \pm \varepsilon_1\varepsilon_2\right),  \\
	&=\Gamma_1\Gamma_2
  	  \begin{cases}
		  2 -(\varepsilon_1 + \varepsilon_2) +
		  \varepsilon_1\varepsilon_2, & \text{+ sign},  \\
		\varepsilon_1 + \varepsilon_2 - \varepsilon_1\varepsilon_2,
		& \text{- sign},
	  \end{cases}  \\
	 &=\Gamma_1\Gamma_2
	  \begin{cases}
		2 -\frac{1}{2}(\Gamma_{1}^{-2}+\Gamma_{2}^{-2})
		+\frac{1}{4}\Gamma_1^{-2}\Gamma_2^{-2},
	 & \text{+ sign},  \\
		\frac{1}{2}(\Gamma_{1}^{-2}+\Gamma_{2}^{-2}) - \frac{1}{4}
		\Gamma_1^{-2}\Gamma_2^{-2},
 	 & \text{- sign},
	  \end{cases} \\
	 &=\begin{cases}
	  2\Gamma_1\Gamma_2
	  -\frac{1}{2}\left(\frac{\Gamma_{2}}{\Gamma_{1}}+\frac{\Gamma_{1}}{\Gamma_2}\right),
		 & \text{+ sign, to  $\mathcal{O}(\Gamma^{-4})$},\\
	  \frac{1}{2}\left(\frac{\Gamma_{2}}{\Gamma_{1}}+\frac{\Gamma_{1}}{\Gamma_2}\right),
		 & \text{- sign, to 
		 $\mathcal{O}(\Gamma^{-4})$}.
	  \end{cases}
\end{split}
\end{equation*}

As a direct consequence, it is found that:

\begin{corollary}

  The sum of the Lorentz factors of \eqref{prop1} satisfy:

\begin{equation}
  \Gamma_{v_2 \oplus v_1} + \Gamma_{v_2 \ominus v_1} = 2\Gamma_1\Gamma_2.
\end{equation}
\end{corollary}

  To compute the square of relations~\eqref{prop1}, note that according
to equation~\eqref{apgamm1}:

\begin{equation*}
 \begin{split}
	\Gamma^2_{v_2 \opm v_1} &=  \Gamma_{1}^2\Gamma_{2}^2 
	  (1 \pm v_2 v_1)^2,  \\
	&=\Gamma_1^2\Gamma_2^2
  	  \begin{cases}
		  (2 -\varepsilon_1 + \varepsilon_2)^2, & \text{+ sign },\\
		  (\varepsilon_1 + \varepsilon_2)^2,     & \text{- sign },
	  \end{cases}  \\
	&\approx\Gamma_1^2\Gamma_2^2
  	  \begin{cases}
		  4(1 -(\varepsilon_1 + \varepsilon_2)), & \text{+ sign},  \\
		  \varepsilon_1^2 +
		  \varepsilon_2^2+2\varepsilon_1\varepsilon_2, & \text{-sign },
	  \end{cases}  \\
	&\approx\Gamma_1^2\Gamma_2^2
  	  \begin{cases}
		  4(1 -(\Gamma_{1}^{-2}+ \Gamma_{2}^{-2}), & \text{+ sign},\\
		  \frac{1}{4}(\Gamma_{1}^{-4}+
		  \Gamma_{2}^{-4})+\frac{1}{2}\Gamma_1^{-2}\Gamma_{2}^{-2},
		  & \text{- sign},
	  \end{cases}  \\
	&\approx
  	  \begin{cases}
		  4\Gamma_1^2\Gamma_2^2, & \text{+ sign},\\
		  \frac{1}{4}\left(\frac{\Gamma_{2}^{2}}{\Gamma_{1}^2}+
		  \frac{\Gamma_{1}^{2}}{\Gamma_{2}^2}\right)+\frac{1}{2},
		  & \text{- sign},
	  \end{cases}
 \end{split}
\end{equation*}

\noindent and so:

\begin{theorem}
  By the same assumptions of Theorem~\ref{teoremita} it follows that:

\begin{align}
	\Gamma^2_{v_2 \oplus v_1} &=  4\Gamma_1^2\Gamma_2^2, \label{a08}\\
	\Gamma^2_{v_2 \ominus v_1} &=\frac{1}{2} +
	\frac{1}{4}\left(\frac{\Gamma_{2}^{2}}{\Gamma_{1}^2}+
	\frac{\Gamma_{1}^{2}}{\Gamma_{2}^2}\right). \label{a09}
\end{align}

\end{theorem}

To conclude this section, consider that one of the two velocities
in \eqref{apgamm1} has any arbitrary value, not necessarily
ultra-relativistic.  In such a case the following Theorem is satisfied:

\begin{theorem}
  If \( \Gamma_1 \gg 1 \) (i.e. when \( v_1 = 1 - 1 / {2\Gamma_{1}^2}
\)) then:

\begin{equation}
  \Gamma^2_{v_1 \opm v_2}=\Gamma_{1}^2\Gamma_{2}^2(1 \pm v_2)^2 \
    \text{for} \ \Gamma_1\gg1,
\label{v2little} 
\end{equation}

\noindent for any value \( v_2 < 1 \).

\end{theorem}

  To proove this theorem note that:
  
\begin{equation*}
 \begin{split}
	\Gamma^2_{v_1 \opm v_2} 
	 &=  \Gamma_{1}^2\Gamma_{2}^2(1 \pm v_1 v_2)^2, \\
	 &=  \Gamma_{1}^2\Gamma_{2}^2\left(1 \pm v_2 \left[ 
	     1-\frac{1}{2\Gamma_{1}^2} \right]\right)^2, \\
	 &=  \Gamma_{1}^2\Gamma_{2}^2\left(1 \pm v_2 \mp 
	     \frac{v_2}{2\Gamma_{1}^2} \right)^2,  \\
	 &=  \Gamma_{1}^2\Gamma_{2}^2\left((1 \pm v_2)^2 \mp 
	     \frac{(1\pm v_2)v_2}{\Gamma_{1}^2} + ... \right).
 \end{split}
\end{equation*}

%%%%%%%%%%%%%%%%%%%%%%%%%%%%%%%%%%%%%%%%%%%%%%%%%%

% Don't change these lines
\bsp	% typesetting comment
\label{lastpage}
\end{document}